\documentclass[]{aastex7}
\usepackage{amsmath}
\usepackage{arydshln}
\usepackage{changepage}
\usepackage{rotating}       
\usepackage{multirow}  
\usepackage{booktabs}       
\usepackage{siunitx}        
\usepackage{tabularx}       
\usepackage{ragged2e}   
\usepackage{threeparttable}

\definecolor{DarkGreen}{rgb}{0.0,0.4,0.0}  
\newcommand{\B}[1]{{\color{blue}{#1}}}
\newcommand{\R}[1]{{\color{red}{#1}}}
\newcommand{\G}[1]{{\color{DarkGreen}{#1}}}
\newcommand{\M}[1]{{\color{magenta}{#1}}}
\newcommand{\C}[1]{{\color{cyan}{#1}}}
\usepackage[normalem]{ulem}
\usepackage{soul}
\usepackage{cancel}

\begin{document}

\title{Waiting Time Distribution of Supra-Arcade Downflows}

\author[gname=Zichao, sname=Yang]{Zichao Yang}
\affiliation{CAS Key Laboratory of Geospace Environment, Department of Geophysics and Planetary Sciences, University of Science and Technology of China, Hefei 230026, People’s Republic of China}
\email{yzc865138921@mail.ustc.edu.cn}  

\author[orcid=0000-0003-4618-4979, gname=Rui, sname=Liu]{Rui Liu} 
\affiliation{CAS Key Laboratory of Geospace Environment, Department of Geophysics and Planetary Sciences, University of Science and Technology of China, Hefei 230026, People’s Republic of China}
\affiliation{Mengcheng National Geophysical Observatory, University of Science and Technology of China, Hefei 230026, People’s Republic of China}
\email{rliu@ustc.edu.cn}

\correspondingauthor{Rui Liu}
\email{rliu@ustc.edu.cn}


\begin{abstract}
    Supra-arcade downflows (SADs) are dark, sunward-moving structures above the arcade of flare loops. Naturally they are considered to be associated with outflows resulting from magnetic reconnection at the vertical current sheet beneath the eruptive structure. Despite the extensive investigations on the SAD properties, the timing information, particularly the waiting time between consecutive SADs, has not been systematically examined. Here, using the 131~{\AA} passband of the Atmospheric Imaging Assembly on-board the Solar Dynamics Observatory, we studied the waiting time distribution (WTD) of SADs in 7 eruptive flares. In six of the 7 flares, the SADs are identified and tracked in previous studies, by two different methods; and in the 7th flare, by our optimized manual method. Based on statistical evaluations, we found that most of the WTDs are best fitted by a power-law function with the slope ranging from 1.7 to 2.4, in accordance with the prediction of non-stationary Poisson processes or self-organized criticality; but often they can also be fitted almost equally well by a log-normal function. These results rule out linear random or quasi-periodic processes to be responsible for the generation of SADs, but suggest that several nonlinear mechanisms be coupled together in the reconnection outflow region to shape the heavy-tailed WTD of SADs.
\end{abstract}



\section{Introduction} \label{sec:1}

Solar flares and coronal mass ejections (CMEs) are the most energetic phenomena in the solar system. It is generally believed that magnetic reconnection, a fundamental mechanism in magnetized plasma, plays a key role in flares and CMEs \cite[]{Lin2004,Forbes2006}. From an edge-on perspective, a plasma sheet observed above the cusp-shaped flare loop is often interpreted to be the heated turbulent plasma surrounding the vertical current sheet in the standard flare model, where the magnetic energy is explosively released and then accelerated energetic particles \cite[]{Lin2007,Ciaravella2008,Liu2013,Reeves2019}. Yet from a face-on perspective, hot but diffuse plasma is observed above the post-fare arcade, known as the supra-arcade fan \cite[SAF;][]{Hanneman2014,reeves2017exploration,Cai2019}, which is often interwoven with spike-like structures \cite[]{Liu&Wang2021}. These features are only visible in EUV filters sensitive to hot plasmas of temperatures at above 10 MK, such as the 131~{\AA} passband (primarily \ion{Fe}{21}; $\log T \mathrm{[K]} = 7.05$) of the Atmospheric Imaging Assembly \cite[AIA;][]{Lemen2012} on the Solar Dynamics Observatory \cite[SDO;][]{Pesnell2012}. 

Dark, tadpole-like voids, known as supra-arcade downflows \cite[SADs;][]{vsvestka1998large,mckenzie1999x,innes2003sumer}, are often observed to move through the SAF toward the post-flare arcade during the early decay phases of flares associated with CMEs. The nature of SADs remains elusive, although they are believed to originate from reconnection outflows or contracting magnetic loops formed via magnetic reconnection \cite[e.g.,][]{savage2012re, Liu2013, mckenzie2013turbulent, Guo2014,shen2022origin, Awasthi2022}.

Previous studies have extensively investigated the kinematic and thermodynamic properties of SADs. Observations show that SADs have a typical lifetime of a few minutes, and their velocities range from 100 to 500 km~s$^{-1}$. With the differential emission measure (DEM) analysis, SADs are found to be cooler than the surrounding fan plasma \cite[]{hanneman2014thermal}, but can potentially heat up local plasma via compression or viscous dissipation as they move through the surrounding medium in a turbulent fashion \cite[]{reeves2017exploration,xue2020thermodynamical,li2021thermodynamic,tan2022statistical}. Furthermore, \cite{samanta2021plasma} proposed that collisions between SADs and post-flare loops can transiently heat the plasma to temperatures of 10–20 MK, and since different SADs collide with the loops at different occasions, the resultant EUV/X-ray emission is enhanced quasi-periodically. In addition, 
the frequency distributions of  SAD parameters including the width, velocity, and lifetime can be well described by log-normal functions \cite[]{xie2022statistical} and compare favorably with the numerical simulation by \cite{shen2022origin}, in which SADs are argued to be self-organized structures formed in the turbulent interface region between the reconnection outflows and the flare arcade. More recently, \cite{Xie2025} argued that SADs are part of a broader turbulent flow system, with anisotropic dynamics that vary with height in the supra-arcade fan. \cite{French2025} reported the first spectroscopic measurements of SADs, revealing distinct Doppler shifts, enhanced nonthermal velocities, and temperatures comparable to the surrounding plasma. These findings highlight the complex nature of SADs.

Despite the above studies on the SAD properties, it is still unknown whether SADs are generated by a quasi-periodic process (e.g., MHD wave or loop oscillations), or an uncorrelated linear random process (e.g., stationary Poisson process), or a nonlinear process with intermittency, clustering, and memory (e.g., SOC, nonstationary Poisson, or turbulence). In this regard, the analysis of waiting times can help distinguish different types of random processes \cite[][Chap 5]{Aschwanden2011}, potentially providing clues to the generation mechanism of SADs. In this paper, we present a statistical study of SAD waiting times with seven limb flares observed by SDO/AIA, in which SADs are identified by different approaches including automated optical flow tracking \cite[]{xie2022statistical}, manual identification \cite[]{tan2022statistical}, and an optimized manual detection method that we develop in this paper. Section \ref{sec:2} describes the AIA observation and our detection method, Section \ref{sec:3} details the statistical procedures and results. Section \ref{sec:4} offers the conclusions.

\section{Event Selection and Detection Methods} \label{sec:2}
The waiting time statistics requires that a flare must have a large number of well-defined SADs. This places a stringent limit on the event selection; e.g., in a sample of 35 limb flares, only 3 have more than 30 SADs \cite[]{Savage&McKenzie2011}. The difficulty also lies in measuring reliably and objectively the characteristics of SADs. In two recent statistical studies on SADs \cite[]{xie2022statistical,tan2022statistical}, the events selected largely satisfy the above requirement, and the start/end times of each identified SAD are provided through either time-lapse animations of 131~{\AA} images in which SADs are marked \cite[]{xie2022statistical} or a table of detailed SAD information \cite[]{tan2022statistical}. We hence derived the waiting times of SADs from the six flares selected by \cite{xie2022statistical} (Case A-F in Table \ref{tab:info}) and from \cite{tan2022statistical} (Case G in Table \ref{tab:info}, the same flare as Case F). SADs in Cases A--F are identified by an automatic recognition algorithm \cite[]{xie2022statistical}, while those in Case G are manually tracked frame by frame \cite[]{tan2022statistical}. In addition, we processed a new flare event occurring at the southeastern limb on 10 February 2024 (Case H in Table \ref{tab:info}). The source region, NOAA AR 13584, was behind the limb during the eruption. The flare starts at 03:04 UT and peaks at 03:54 UT; afterwards numerous episodes of SADs are continually observed for more than two hours. So in total we have data from seven flare events, as shown in Table \ref{tab:info}. 

\begin{figure}[!ht]
    \centering
    \includegraphics[width=0.7\linewidth]{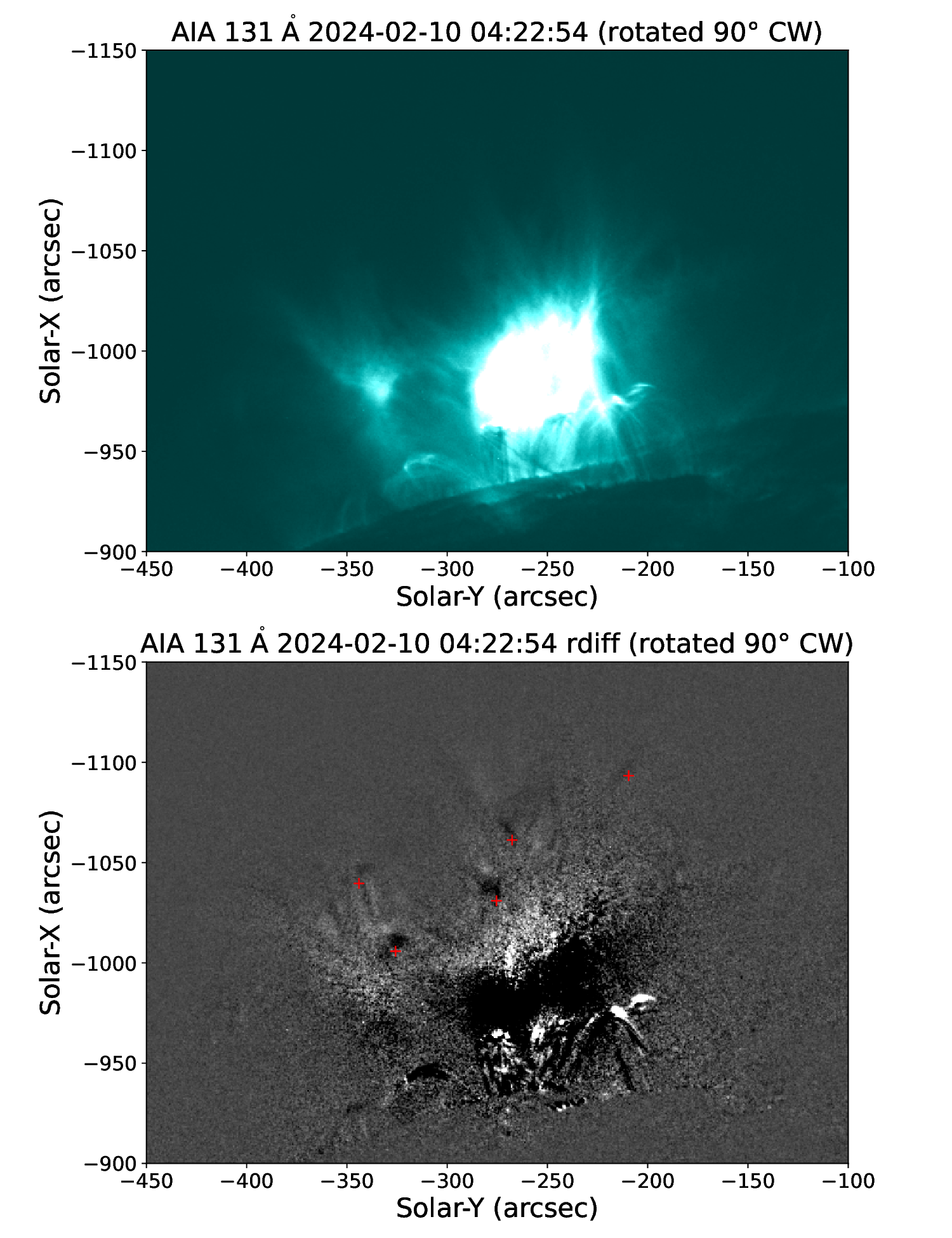}
    \caption{SADs observed in Case H. An AIA 131~{\AA} image  is shown in the upper panel, and the corresponding 1-min running difference image in the lower panel, where SADs are marked by red `+' signs. Note that the images have been rotated 90 degree clockwise. An animation AIA 131~{\AA} and corresponding running difference images is available online.}
    \label{fig:Case H}
\end{figure}

To detect SADs, we utilized running difference images at 1-min cadence to enhance the dynamic motions and the characteristic reduced brightness of SADs in the AIA 131~{\AA} passband. We then generated animation sequences with a fixed-length sliding window: each sequence consists of 10-frame stacks of both original and difference images, but the subsequent sequence is created by advancing the window position by only one frame, so as to preserve the original temporal resolution. With the animation sequences, we manually marked the SAD locations, and tracked their temporal evolution to determine the start and end times of each SAD. By combining difference imaging with sliding window animation, our approach enhances SAD features while taking advantage of human motion perception, so as to achieve reliable SAD identification and tracking, despite complex dynamics and weak emission in the supra-arcade region. Figure \ref{fig:Case H} shows the original and running difference image, in which the identified SADs are marked by red plus signs. Recording and sorting the start and end time of each individual SADs, we obtained the waiting times in the flare. The waiting time is defined by the time interval between successive SAD start/end time. A caveat to keep in mind is that the first/last visible appearance of an SAD does not directly correspond to its true physical formation/demise time, but marks the time range during which the SAD remains detectable under observational and instrumental limitations. Additionally, due to the limited temporal resolution, waiting times between those SADs that appear/disappear in the same frame are simply ignored in the statistics. 

\begin{table*}[htb]
\begin{adjustwidth}{-3cm}{-3cm}
\centering \small
\begin{threeparttable}
\caption{\label{tab:info}List of the flares and SAD cases under investigation}
\begin{tabular}{cccccc}
\toprule
Date & Flare Class & Time(UT) & Case & Sample Size \\
\midrule
2011 Oct 22 & M1.3 & 12:00–13:29 & A & 69 \\
2015 Jun 18 & M1.3 & 01:20–02:52 & B & 71 \\
2012 Jul 17 & M1.7 & 15:34–17:16 & C & 77 \\
2012 Jan 16 & C6.5 & 03:30–05:50 & D & 36 \\
2012 Jan 27 & X1.7 & 19:02–21:59 & E & 133 \\
\multirow{2}{*}{2013 May 22} & \multirow{2}{*}{M5.0}& 
    13:28–16:20 & F & 173 \\
    &&13:29-16:45 & G & 69 \\
2024 Feb 10 & M3.4 & 03:50-06:04 & H & 128 \\
\bottomrule
\end{tabular}
\end{threeparttable}
\end{adjustwidth}
\tablecomments{
The column `Sample Size' refers to the count of valid waiting time intervals, not the quantity of SADs.
} 
\end{table*}

\section{Data Analysis and Result} \label{sec:3}

In this section, we present the the waiting time distributions (WTDs) of SADs identified in the 7 flares under investigation. We then fit the observed WTDs with candidate models including log-normal, exponential, and power-law distributions, and apply various statistical criteria to evaluate the goodness of fit in order to discriminate the models. 

\subsection{Waiting time distributions and fittings} \label{sec:3.1}

\begin{figure}[ht]
    \centering
    \includegraphics[width=\linewidth]{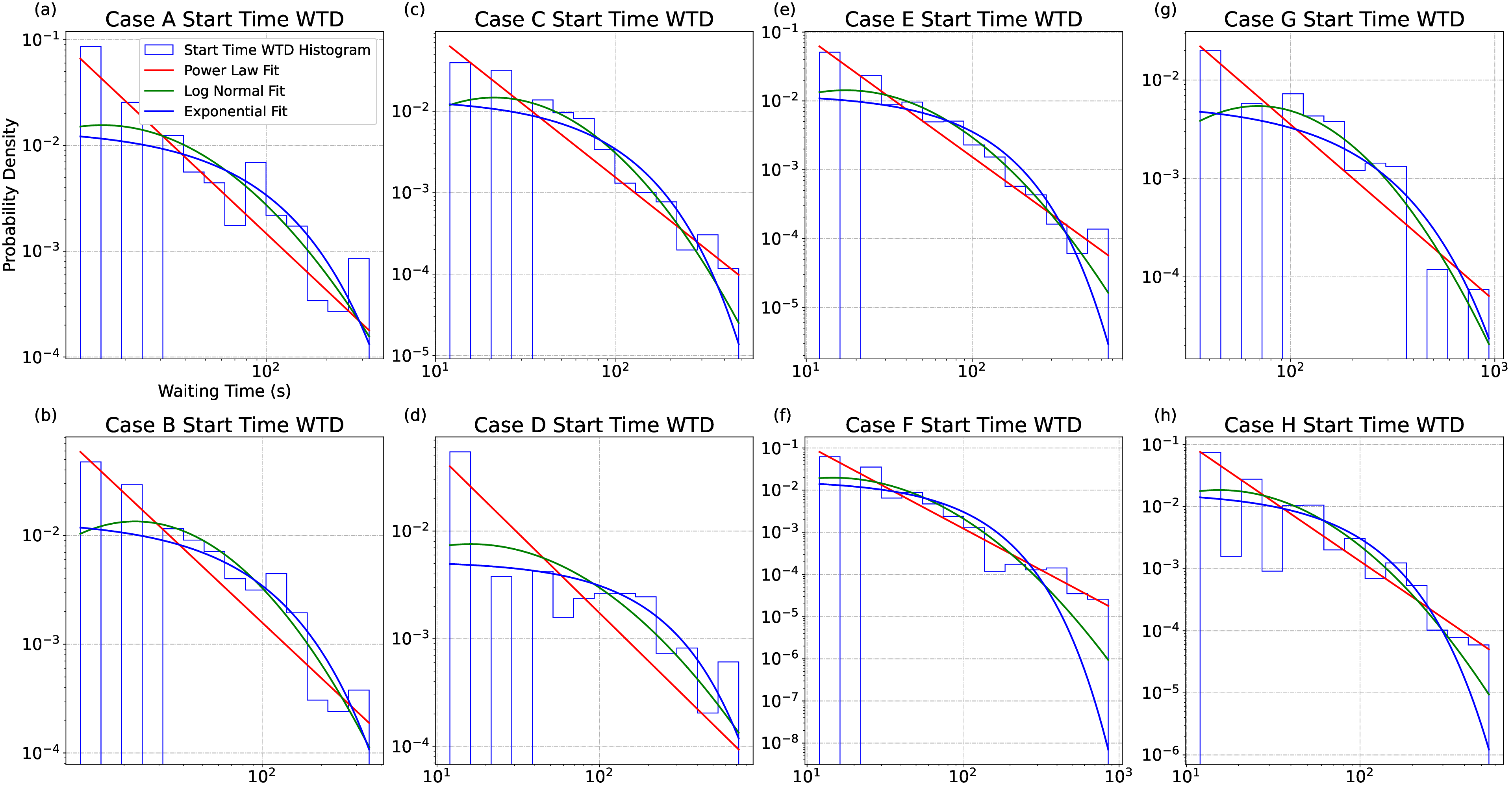}
    \caption{WTDs derived from the start time of SADs. The histograms are normalized to show the probability density, and fitted with the MLE method for power-law (red), log-normal (green), and exponential (blue) functions. The fitting parameters are shown in Table \ref{tab:startresults}.} 
    \label{fig:fitting_start}
\end{figure}

The waiting times of the identified SADs are modeled using three different probability distributions, namely, log-normal, exponential, and power-law distributions:  
\begin{gather}
    p(x;\sigma,\mu)=\frac{1}{x\sigma \sqrt{2\pi}}e^{-\frac{(\ln x-\mu)^2}{2\sigma ^2}}; \label{eq:lognormal}\\ 
    p(x;\lambda)=\lambda e^{-\lambda x} \label{eq:exponential};\\
    \begin{aligned}[t]
        p(x;\alpha,x_{\min})=\frac{\alpha-1}{x_{\min}} \left(\frac{x}{x_{\min}}\right)^{-\alpha}.       
    \end{aligned}
    \label{eq:powerlaw}
\end{gather}
The log-normal distribution depends on the parameters $\mu$ (logarithmic mean) and $\sigma$ (logarithmic standard deviation), the exponential distribution is governed by the mean event rate $\lambda$, and the power-law distribution by the exponent $\alpha$. Parametric estimation for all three candidate distributions was performed through maximum likelihood estimation (MLE; Table \ref{tab:startresults}). For example, the MLE estimated $\alpha$ for power-law distributions is given as follows, 
\begin{equation}
    \hat{\alpha} = 1+n\left[\sum_{i=1}^{n}\ln{\frac{x_i}{x_{\min}}}\right]^{-1},
\end{equation}
which is sensitive to the choice of ${x_{\min}}$, the lower bound of the power-law range. Here ${x_{\min}}$ is naturally specified as the minimum waiting time of each data set (36~s for Case G and 12~s for other cases). Figure \ref{fig:fitting_start} shows the WTDs in histograms and the fitting curves of three candidate distributions in different colors. 

We also applied the survival distribution function (SDF) method to estimate the power-law slope \cite[Section 2.4.3]{Survival2018}. For a random variable $x$ with the probability density function (pdf), $p(x)$, the survival function is defined as 
\begin{equation}
    sf(x) = P(X\ge x)=\int_{x}^{\infty}p(x')dx'. \label{eq:sf-def}
\end{equation}
The SDF method has advantages when the number of data points available is small, because the survival function can be trivially but accurately constructed from the raw waiting time data by a simple reordering without any binning \cite[Equation 2.96--2.101]{Survival2018}. If the pdf is in the form of power law (Eq.~\ref{eq:powerlaw}), then the survival function can be given analytically as follows,
\begin{gather}
    sf(x) = \left(\frac{x}{x_{\min}}\right)^{1-\alpha}.
    \label{eq:sf-powerlaw}
\end{gather}
Hence by fitting the reconstructed survival function with Eq.~\ref{eq:sf-powerlaw}, we can get an estimate of $\alpha$. This is done by a linear regression fit on a log-log scale to prioritize the fitting of the power-law tail. In contrast, fitting the data with Eq.~\ref{eq:sf-powerlaw} directly on the linear scale would give more weight to short waiting times, resulting in an underestimation of $\alpha$. The power-law indices obtained through the SDF method ($\alpha_\mathrm{SDF}$) are shown in Table \ref{tab:startresults}, along with those obtained from the MLE method ($\alpha_\mathrm{MLE}$). It can be seen that all of $\alpha_\mathrm{MLE}$ are slightly below 2 whereas most of $\alpha_\mathrm{SDF}$ are slightly above 2. Such discrepancies can be taken as the uncertainty resulting from the small number of identified SADs. We must also keep in mind that observational limitations may introduce systematic biases; e.g., if the observation fails to detect small-scale SADs due to the limited spatio-temporal resolution, it would result in the under-sampling of short waiting times, therefore yielding a flatter distribution.

\subsection{Statistical Evaluation} \label{sec:3.2}

From Figure~\ref{fig:fitting_start}, one may get an impression that the data aligns more closely with a power-law distribution. To determine the most suitable model objectively, we employed the following statistical tests and criteria, including the likelihood ratio test \cite[LRT;][Appendix C]{clauset2009power}, Akaike information criterion \cite[AIC;][]{AIC1974}, corrected Akaike information criterion \cite[AICc;][]{AICc1978}, and Bayesian information criterion \cite[BIC;][]{BIC1978}. These tests enable quantitative evaluation of model performance through both hypothesis-testing (LRT) and information-theoretic (AIC/AICc/BIC) frameworks. First, we performed pairwise LRT comparisons among the three candidate distributions. LRT yields $p$- and $R$- values: a smaller $p$-value indicates higher confidence in the corresponding $R$-value; a positive (negative) $R$-value suggests a preference for the first (second) distribution in a pairwise comparison. Second, we compute the AIC, AICc, and BIC values for each fitted distribution. These criteria impose progressively stronger penalties (AIC $<$ AICc $<$ BIC) on model complexity (e.g., the number of parameters), and the more favored distribution is assigned smaller values. 


The testing results of both approaches as shown in Table \ref{tab:startresults} demonstrate that Cases A, E, F, and H favor the power-law distribution, that Cases B and G favors the log-normal distribution, and that the remaining cases are ambiguous (Cases C \& D). 

Specifically, the LRT indicates that the power-law distribution provides the best fitting for Cases A, E, F, and H, with log-normal distribution ranking the 2nd and exponential distribution the 3rd. In contrast, the LRT suggests the log-normal distribution as the optimal fit for Case G, with exponential distribution ranking the second and power-law the third. Cases B and C demonstrate that the log-normal distribution outperforms the exponential distribution; however, the remaining pairwise comparisons (e.g., log-normal vs. power-law, exponential vs. power-law) yield statistically insignificant results ($p>0.1$). Similarly, none of the three pairwise comparisons in Case D are statistically significant. 

The three model selection criteria (AIC, AICc, BIC) yield consistent conclusions: the power-law distribution is identified as the optimal model for Cases A, C, E, F, and H. In Case C, the power-law only marginally outperforms the log-normal. Cases B and G show a slight preference for the log-normal distribution over power-law. Case D selects the exponential distribution as the best fit.

\begin{sidewaystable}[p]
\centering
\caption{Fitting results and statistical tests for the WTDs of SADs}
\label{tab:startresults}
\small
\setlength{\tabcolsep}{4pt} 
\begin{tabular*}{\textheight}{@{\extracolsep{\fill}}lcccccccccc@{}}
\toprule
& & & \multicolumn{8}{c}{Dataset} \\ 
\cmidrule{4-11}
& & & A & B & C & D & E & F & G & H \\ 
\midrule

\multirow{5}{*}{Parameters}  
& \multirow{2}{*}{Power-law}
            & $\alpha_\mathrm{MLE}$ & 1.795 & 1.705 & 1.751 & 1.480 & 1.745 & 1.973 & 1.793 & 1.913 \\
            & & $\alpha_\mathrm{SDF}$ & 2.187 & 2.311 & 2.227 & 1.773 & 2.210 & 2.234 & 2.395 & 2.210 \\ 
\addlinespace[1ex]
& \multirow{2}{*}{Log-norm} 
        & $\sigma$ & 1.004 & 0.883 & 0.875 & 1.338 & 0.990 & 0.911 & 0.781 & 0.919 \\
        &  & $\mu$ & 3.743 & 3.903 & 3.816 & 4.568 & 3.828 & 3.513 & 4.845 & 3.580 \\
\addlinespace[1ex]
& Exponential & $\lambda$ & \num{1.45e-2} & \num{1.40e-2} & \num{1.45e-2} & \num{5.27e-3} & \num{1.27e-2} & \num{1.73e-2} & \num{5.90e-3} & \num{1.73e-2} \\
\midrule

\multirow{6}{*}{LRT} 
& \multirow{2}{*}{Log-norm vs. Power-law}
 & $R$ & \textbf{-13.343} & 3.886 & 0.072 & 0.857 & \textbf{-15.113} & \textbf{-51.634} & 4.171 & \textbf{-31.130} \\
& & $p$ & 0.057 & 0.586 & 0.992 & 0.881 & 0.108 & \num{1.41e-6} & 0.587 & \num{6.25e-4} \\
\addlinespace[1ex]
& \multirow{2}{*}{Log-norm vs. Exponential}
 & $R$ & \textbf{4.723} & \textbf{5.172} & \textbf{10.294} & -1.184 & \textbf{17.418} & \textbf{38.142} & \textbf{8.054} & \textbf{18.130} \\
& & $p$ & \num{7.57e-3} & \num{2.98e-3} & \num{4.23e-7} & 0.536 & \num{5.70e-7} & \num{1.20e-9} & \num{8.14e-5} & \num{6.30e-8} \\
\addlinespace[1ex]
& \multirow{2}{*}{Exponential vs. Power-law}
 & $R$ & \textbf{-18.067} & -1.286 & -10.222 & 2.041 & \textbf{-32.531} & \textbf{-89.776} & -3.883 & \textbf{-49.261} \\
& & $p$ & 0.029 & 0.853 & 0.149 & 0.784 & \num{4.58e-3} & \num{2.46e-9} & 0.559 & \num{1.01e-5} \\
\midrule

\multirow{3}{*}{AIC}
& Power-law & & \textbf{688.26} & 747.71 & \textbf{787.70} & 455.74 & \textbf{1364.73} & \textbf{1572.88} & 840.62 & \textbf{1197.88} \\
& Log-norm & & 716.95 & \textbf{741.94} & 789.55 & 456.03 & 1396.95 & 1678.15 & \textbf{834.28} & 1262.14 \\
& Exponential & & 724.39 & 750.28 & 808.14 & \textbf{451.66} & 1429.79 & 1752.43 & 848.39 & 1296.40 \\
\midrule

\multirow{3}{*}{AICc}
& Power-law & & \textbf{688.32} & 747.77 & \textbf{787.75} & 455.86 & \textbf{1364.76} & \textbf{1572.90} & 840.68 & \textbf{1197.91} \\
& Log-norm & & 717.13 & \textbf{742.12} & 789.71 & 456.39 & 1397.05 & 1678.22 & \textbf{834.46} & 1262.24 \\
& Exponential & & 724.45 & 750.34 & 808.19 & \textbf{451.78} & 1429.82 & 1752.45 & 848.45 & 1296.43 \\
\midrule

\multirow{3}{*}{BIC}
& Power-law & & \textbf{690.49} & 749.97 & \textbf{790.04} & 457.32 & \textbf{1367.62} & \textbf{1576.03} & 842.86 & \textbf{1200.73} \\
& Log-norm & & 721.41 & \textbf{746.46} & 794.24 & 459.19 & 1402.74 & 1684.45 & \textbf{838.75} & 1267.85 \\
& Exponential & & 726.63 & 752.55 & 810.48 & \textbf{453.24} & 1432.68 & 1755.58 & 850.62 & 1299.25 \\

\bottomrule
\end{tabular*}
\tablecomments{The waiting times are derived from the start times of SADs. For LRT, comparisons yielding significant results ($p < 0.1$) are shown in boldface. Note that positive $R$ favors Model \#1, and negative $R$ favors Model \#2. For AIC/AICc/BIC tests, the lowest value that indicates the best fit is also shown in boldface.}

\end{sidewaystable}


Among the 7 flares, Case D seems to be an outlier: the LRT fails to produce significant results ($p>0.1$), but the model selection criteria favor the exponential distribution. The SAF in Case D exhibits pronounced projection effects due to the inclined viewing geometry of SDO/AIA \cite[][Appendix B.4 and Figure 22]{xie2022statistical}. Consequently, some SADs are obscured, only 39 SADs are identified. Although the number still falls within the upper range among limb flares \cite[e.g.,][]{Savage&McKenzie2011}, it is significantly fewer than other cases. The small sample size may compromise both the quality and reliability of the statistics for Case D. 

The above results are derived from the start time of SADs, we also analyze the results from the end time of SADs and find a largely similar result, with power-law being favored in five of the eight cases (Cases A, C, F, G, and H); Cases B and E favors log-normal distribution, while Case D prefers exponential distribution. Notably, Cases E and G switch their optimal models between start and end time statistics. 

Cases F and G provide an opportunity to compare different tracking method on the same event. The automated method by \cite{xie2022statistical} identified 230 SADs, in comparison to 81 manually identified by \cite{tan2022statistical}. Some SADs might be missed by manual identification since the object of \cite{tan2022statistical} is to study SAD trajectories; naturally SADs that only appear in one or two frames tend to be ignored. The preference for log-normal in Case G may result from the undersampling of short waiting times. On the other hand, the automated method inevitably produces false positives. However, despite the large difference in the number of SADs, both cases yield heavy-tailed WTDs; and for the end time result, Case G favors power-law. It is remarkable that Case H, in which SADs are manually identified by us, also favors power-law. 

Therefore, the cases under investigation generally support the power-law distribution in characterizing the WTDs of SADs when we exclude Case D as an outlier. However, in several cases, the log-normal model provides a comparable fit and remains a plausible alternative. In those cases favoring log-normal distribution, the power-law model remains plausible to account for the tail of the WTDs. Overall, both models consistently reflect the heavy-tailed and non exponential nature of the WTDs.


\section{Discussion} \label{sec:discuss}

Here we put the observed WTDs of SADs in the perspective of the mechanisms capable of generating power-law distributions, namely, self-organized criticality (SOC), non-stationary Poisson process, and MHD turbulence. The implications of the log-normal distribution, which fits the WTDs almost equally well, are also discussed in the context of recent simulations on SADs. 

The fractal-diffusive SOC model proposed by \cite{aschwanden2014macroscopic} provides a macroscopic unified theoretical framework for power-law distributions in astrophysical phenomena. This model derives the power-law form of WTDs based on the assumption of geometric probability conjectures, i.e., $N(L)dL \propto L^{-d}dL$, where $d$ is the Euclidean space dimension, and on the assumption of  a scaling relation between spatial scale $L$ and temporal scale $T$, $L \propto \kappa T^{\beta/2}$, resulting from diffusive processes. For three-dimensional space ($d = 3$) and classical diffusion ($\beta = 1$), the theoretical prediction for the power-law index of waiting time distributions \cite[][Eq.~19]{aschwanden2014macroscopic} is:
\begin{equation}
    \alpha_{\Delta t} = 1 + (d-1) \times \frac{\beta}{2} = 2, 
    \label{eq:SOC_formula_1}
\end{equation}
which requires that the upper limit of the inertial range of time scales $[T_1, T_2]$ is shorter than the average waiting time, $T_2 < \langle \Delta t \rangle$, so that the avalanches occur independently without any time overlapping between each other. In the opposite case, when $T_2 > \langle \Delta t \rangle$, we begin to see time-overlapping events, a situation referred to as ``event pileup'', and the power-law slope is modified as follows \cite[][Eq.~33]{aschwanden2014macroscopic},
\begin{equation}
    \alpha_{\Delta t}^\mathrm{pileup} = \alpha_{\Delta t} \cdot \frac{\log T_2}{\log \langle \Delta t \rangle}. 
\label{eq:SOC_formula_2}
\end{equation}

The waiting-time statistics in complex systems can also be modeled by a non-stationary Poisson process. The WTD of a classical Poisson process follows an exponential function with a constant occurrence rate $\lambda$. In contrast, the event occurrence rate $\lambda(t)$ varies with time in a non-stationary Poisson process, which accounts for high event rate during active phases and significantly reduced rate during quiescent periods. Such temporal variability typically leads to power-law tails. Formally, the WTD for a non-stationary Poisson process is derived by integrating over all possible rates weighted by their temporal prevalence \cite[][Eq.~3]{wheatland2000origin}:
\begin{equation}
    P(\Delta t) = \frac{1}{\lambda_0}\int_{0}^{\infty}f(\lambda)\lambda^{2}e^{-\lambda\Delta t}d\lambda,
    \label{eq:non-stationary_Poisson_1}
\end{equation}
where $f(\lambda)$ represent the time-dependent rate distribution, i.e., $f(\lambda)d\lambda$ is the fraction of time for the occurrence rate in the range $(\lambda,\lambda+d\lambda)$, with $\lambda_0$ indicating the mean rate. \cite{aschwanden2010reconciliation} derived the form of $P(\Delta t)$ in five different situations and found that smoothly varying occurrence rates predominantly yield approximate power-law distributions at large waiting times, with a power-law index in the range 2--3 with an upper bound 3 \cite[]{aschwanden2010reconciliation}. With a flare rate varying as $f(\lambda)\propto \lambda^{-1}\exp{(-\lambda/\lambda_0)}$, \cite{aschwanden2010reconciliation} proposed a unified WTD, $N(\Delta t)\propto \lambda_0(1+\lambda_0\Delta t)^{-2}$, which has a power-law slope of 2 at large waiting times but flattens out at short waiting times, to account for the observed WTDs of flares. \cite{Aschwanden2021poisson} parameterized the event rate function with an exponent $p$ so that $\lambda(t)\propto t^p$, which yields a power-law slope $\alpha=2+1/p$ in the range of $[2, 2.5]$ in the nonlinear regime ($p\ge2$).

Previous studies also invoked turbulence to explain power-law-like WTDs. \cite{boffetta1999power} demonstrated that through a magneto-hydrodynamic (MHD) shell model simulation, turbulent intermittency can generate power-law WTDs with a slope of $\sim\,$2.7, close to the experiment result that gives a power-law slope of $\sim\,$2.4. Compared to the self-similar feature of the SOC model, the intermittency in turbulence shows a more chaotic nature, linking to the nonlinear destabilization of laminar phases and reorganization of vortex structures, which fundamentally differs from the exponential decay predicted by SOC (for the quiet phase). \cite{lepreti2001solar} analyzed waiting times using the GOES flare list from 1975 to 1999 and found that the observed WTD can be reproduced by  L\'{e}vy statistics  with an power-law index of $\sim\,$2.38. They also used a local Poisson hypothesis test to prove the unapplicability of a time-varying Poisson process. \cite{grigolini2002diffusion} introduced a diffusion entropy method to derive a power-law index of $\sim\,$2.14, supporting the hypothesis of L\'{e}vy statistics. 

Our analysis reveals that $\alpha_{\Delta t} \approx 1.7-2.0$ for SADs, but the true slope should be higher because of the under-sampling of short waiting times due to limited spatio-temporal resolution. In a pile-up situation like SADs, the standard 3D SOC model also predicts a higher power-law slope, i.e., $\alpha_{\Delta t} > 2$ (Eq.~\ref{eq:SOC_formula_2}), which favors the power-law slope given by the SDF method (\S\ref{sec:3.1} and Table~\ref{tab:startresults}). Alternatively, the power-law distribution may result from non-stationary Poisson processes \cite[$2\le\alpha_{\Delta t} \le 2.5$;][]{Aschwanden2021poisson} through an intermittent energy release process \cite[e.g., Case 5 in Figure 1 of][]{aschwanden2010reconciliation}. The model of turbulence shares a similar nature of intermittency. It is conjectured that turbulent cascades in reconnecting current sheets fragment the sheet into plasmoids, producing L\'{e}vy-like bursts of energy release \cite[]{lepreti2001solar,boffetta1999power,skender2010instability}. 
Thus, the power-law slope of WTDs might point to a hybrid generation mechanism for SADs: an SOC-like criticality governs the macro-scale energy storage and avalanche triggering, while turbulent intermittency introduces spatio-temporal correlations at smaller scales.

However, both power-law and log-normal are plausible models for the heavy-tailed behavior in complex systems. It is challenging to discriminate between the two models, due to the intrinsic ambiguity that arises from fitting finite heavy-tailed distributions \cite[]{willinger2004}. The log-normal distribution indeed provides comparably good fits for several cases under investigation (\S~\ref{sec:3.2}). As discussed in \cite{xie2022statistical}, a log-normal distribution may arise from a multiplicative stochastic process, involving both random fluctuations and instabilities. In their numerical simulation,  SADs are formed in the turbulent interface region, due to a mixture of Rayleigh–Taylor and Richtmyer–Meshkov instabilities, which is not in contradiction with turbulence driven power-law behavior. Thus, the heavy-tailed WTDs may reflect the underlying mechanism that releases energy intermittently in a turbulent environment.

\section{Summary \& Conclusion} \label{sec:4}

In this study, we analyzed the WTDs of SADs observed in seven flare events. Start (or end) times of SADs in Case A-F are given by \cite{xie2022statistical} through automated optical flow tracking, while those in Case G are given by \cite{tan2022statistical} through manual tracking. We introduced an additional case (Case H) and developed a visual recognition routine to enhance objectivity in the SAD identification. With this routine, we marked SADs in motion pictures created through a sliding window method, which preserves the original temporal resolution and enhances the motional features of SADs using running difference images. Then the waiting times are calculated as the intervals between the start (or end) times of consecutive SADs in each case.

We tested three candidate distributions (power-law, log-normal, and exponential) for the WTDs. Statistical evaluations based on likelihood ratio and information criteria (AIC/AICc/BIC) reveal that the WTD for most cases can be well fitted by either a power-law or a log-normal distribution, with a power-law slope ranging from 1.7 to 2.4. Only one case with a small number of SADs favors the exponential distribution. This consistency across different detection methods and flare events suggests that the observed heavy-tailed WTD is intrinsic to the SAD generation and propagation processes. \cite{xie2022statistical} also found that SAD kinetic parameters (e.g., width, velocities, lifetime) follow log-normal statistics, consistent with 3D MHD simulations of turbulent current sheets \cite[]{shen2022origin}. In their simulations, SADs form through Rayleigh–Taylor and Richtmyer–Meshkov instabilities in the interface region, where turbulent fluctuations drive the log-normal WTD. 

Thus, the heavy-tailed nature of the observed WTDs clearly distinguishes themselves from exponential distributions as expected from linear random processes, and from Gaussian-like distributions as expected from quasi-periodic processes, but suggests nonlinear random processes with intermittence, clustering, and memory. We speculate that the generation and motions of SADs may involve multiple mechanisms that are coupled together. At larger scales, the energy storage in the coronal magnetic field and sudden release in the current sheet may be conditioned by SOC. In this framework, energy accumulates gradually until reaching a critical threshold, triggering avalanche-like reconnection events. At smaller scales, turbulent processes dominate, fragmenting the current sheet into plasmoids and driving irregular bursts of energy release. This multi-scale interplay explains the observed power-law distribution in waiting times, suggesting the scale-invariant nature of the flare energy release process. Additional complexity is introduced by the log-normal distribution as revealed in some cases, which could originate from instabilities associated with the turbulent plasma environment through which the SADs propagate. Together, these processes shape the heavy-tailed WTD of SADs.

\begin{acknowledgments}
This work was supported by the National Key R\&D Program of China (2022YFF0503002), the Strategic Priority Program of the Chinese Academy of Sciences (XDB0560102), and the National Natural Science Foundation of China (NSFC; 42274204, 12373064, 42188101, 11925302). Z.Y. thanks Dr. Xiaoyan Xie for helpful discussions.
\end{acknowledgments}

\bibliography{ref}

\begin{thebibliography}{}
\expandafter\ifx\csname natexlab\endcsname\relax\def\natexlab#1{#1}\fi
\providecommand{\url}[1]{\href{#1}{#1}}
\providecommand{\dodoi}[1]{doi:~\href{http://doi.org/#1}{\nolinkurl{#1}}}
\providecommand{\doeprint}[1]{\href{http://ascl.net/#1}{\nolinkurl{http://ascl.net/#1}}}
\providecommand{\doarXiv}[1]{\href{https://arxiv.org/abs/#1}{\nolinkurl{https://arxiv.org/abs/#1}}}

\bibitem[{H. {Akaike}(1974){Akaike}}]{AIC1974}
{Akaike}, H. 1974, \bibinfo{title}{{A New Look at the Statistical Model
  Identification},} IEEE Transactions on Automatic Control, 19, 716

\bibitem[{M.~J. {Aschwanden}(2011){Aschwanden}}]{Aschwanden2011}
{Aschwanden}, M.~J. 2011, {Self-Organized Criticality in Astrophysics}
  (Springer Berlin, Heidelberg), \dodoi{10.1007/978-3-642-15001-2}

\bibitem[{M.~J. {Aschwanden}(2014){Aschwanden}}]{aschwanden2014macroscopic}
{Aschwanden}, M.~J. 2014, \bibinfo{title}{{A Macroscopic Description of a
  Generalized Self-organized Criticality System: Astrophysical Applications},}
  \apj, 782, 54, \dodoi{10.1088/0004-637X/782/1/54}

\bibitem[{M.~J. {Aschwanden} {et~al.}(2021){Aschwanden}, {Johnson}, \&
  {Nurhan}}]{Aschwanden2021poisson}
{Aschwanden}, M.~J., {Johnson}, J.~R., \& {Nurhan}, Y.~I. 2021,
  \bibinfo{title}{{The Poissonian Origin of Power Laws in Solar Flare Waiting
  Time Distributions},} \apj, 921, 166, \dodoi{10.3847/1538-4357/ac19a9}

\bibitem[{M.~J. {Aschwanden} \& J.~M. {McTiernan}(2010){Aschwanden} \&
  {McTiernan}}]{aschwanden2010reconciliation}
{Aschwanden}, M.~J., \& {McTiernan}, J.~M. 2010,
  \bibinfo{title}{{Reconciliation of Waiting Time Statistics of Solar Flares
  Observed in Hard X-rays},} \apj, 717, 683,
  \dodoi{10.1088/0004-637X/717/2/683}

\bibitem[{A.~K. {Awasthi} {et~al.}(2022){Awasthi}, {Liu}, \&
  {Gou}}]{Awasthi2022}
{Awasthi}, A.~K., {Liu}, R., \& {Gou}, T. 2022, \bibinfo{title}{{Effects of
  Supra-arcade Downflows Interacting with the Postflare Arcade},} \apj, 941,
  158, \dodoi{10.3847/1538-4357/aca3a8}

\bibitem[{G. {Boffetta} {et~al.}(1999){Boffetta}, {Carbone}, {Giuliani},
  {Veltri}, \& {Vulpiani}}]{boffetta1999power}
{Boffetta}, G., {Carbone}, V., {Giuliani}, P., {Veltri}, P., \& {Vulpiani}, A.
  1999, \bibinfo{title}{{Power Laws in Solar Flares: Self-Organized Criticality
  or Turbulence?},} \prl, 83, 4662, \dodoi{10.1103/PhysRevLett.83.4662}

\bibitem[{Q. {Cai} {et~al.}(2019){Cai}, {Shen}, {Raymond}, {Mei}, {Warmuth},
  {Roussev}, \& {Lin}}]{Cai2019}
{Cai}, Q., {Shen}, C., {Raymond}, J.~C., {et~al.} 2019,
  \bibinfo{title}{{Investigations of a supra-arcade fan and termination shock
  above the top of the flare-loop system of the 2017 September 10 event},}
  \mnras, 489, 3183, \dodoi{10.1093/mnras/stz2167}

\bibitem[{A. {Ciaravella} \& J.~C. {Raymond}(2008){Ciaravella} \&
  {Raymond}}]{Ciaravella2008}
{Ciaravella}, A., \& {Raymond}, J.~C. 2008, \bibinfo{title}{{The Current Sheet
  Associated with the 2003 November 4 Coronal Mass Ejection: Density,
  Temperature, Thickness, and Line Width},} \apj, 686, 1372,
  \dodoi{10.1086/590655}

\bibitem[{A. {Clauset} {et~al.}(2009){Clauset}, {Shalizi}, \&
  {Newman}}]{clauset2009power}
{Clauset}, A., {Shalizi}, C.~R., \& {Newman}, M.~E.~J. 2009,
  \bibinfo{title}{{Power-Law Distributions in Empirical Data},} SIAM Review,
  51, 661, \dodoi{10.1137/070710111}

\bibitem[{T.~G. {Forbes} {et~al.}(2006){Forbes}, {Linker}, {Chen}, {Cid},
  {K{\'o}ta}, {Lee}, {Mann}, {Miki{\'c}}, {Potgieter}, {Schmidt}, {Siscoe},
  {Vainio}, {Antiochos}, \& {Riley}}]{Forbes2006}
{Forbes}, T.~G., {Linker}, J.~A., {Chen}, J., {et~al.} 2006,
  \bibinfo{title}{{CME Theory and Models},} \ssr, 123, 251,
  \dodoi{10.1007/s11214-006-9019-8}

\bibitem[{R.~J. {French} {et~al.}(2025){French}, {Kazachenko}, {Mihailescu}, \&
  {Reeves}}]{French2025}
{French}, R.~J., {Kazachenko}, M.~D., {Mihailescu}, T., \& {Reeves}, K.~K.
  2025, \bibinfo{title}{{Spectroscopic Observations of Supra-arcade
  Downflows},} \apjl, 986, L16, \dodoi{10.3847/2041-8213/adde59}

\bibitem[{P. {Grigolini} {et~al.}(2002){Grigolini}, {Leddon}, \&
  {Scafetta}}]{grigolini2002diffusion}
{Grigolini}, P., {Leddon}, D., \& {Scafetta}, N. 2002,
  \bibinfo{title}{{Diffusion entropy and waiting time statistics of hard-x-ray
  solar flares},} \pre, 65, 046203, \dodoi{10.1103/PhysRevE.65.046203}

\bibitem[{L.~J. {Guo} {et~al.}(2014){Guo}, {Huang}, {Bhattacharjee}, \&
  {Innes}}]{Guo2014}
{Guo}, L.~J., {Huang}, Y.~M., {Bhattacharjee}, A., \& {Innes}, D.~E. 2014,
  \bibinfo{title}{{Rayleigh-Taylor Type Instabilities in the Reconnection
  Exhaust Jet as a Mechanism for Supra-arcade Downflows in the Sun},} \apjl,
  796, L29, \dodoi{10.1088/2041-8205/796/2/L29}

\bibitem[{W.~J. {Hanneman} \& K.~K. {Reeves}(2014{\natexlab{a}}){Hanneman} \&
  {Reeves}}]{Hanneman2014}
{Hanneman}, W.~J., \& {Reeves}, K.~K. 2014{\natexlab{a}},
  \bibinfo{title}{{Thermal Structure of Current Sheets and Supra-arcade
  Downflows in the Solar Corona},} \apj, 786, 95,
  \dodoi{10.1088/0004-637X/786/2/95}

\bibitem[{W.~J. {Hanneman} \& K.~K. {Reeves}(2014{\natexlab{b}}){Hanneman} \&
  {Reeves}}]{hanneman2014thermal}
{Hanneman}, W.~J., \& {Reeves}, K.~K. 2014{\natexlab{b}},
  \bibinfo{title}{{Thermal Structure of Current Sheets and Supra-arcade
  Downflows in the Solar Corona},} \apj, 786, 95,
  \dodoi{10.1088/0004-637X/786/2/95}

\bibitem[{D.~E. {Innes} {et~al.}(2003){Innes}, {McKenzie}, \&
  {Wang}}]{innes2003sumer}
{Innes}, D.~E., {McKenzie}, D.~E., \& {Wang}, T. 2003, \bibinfo{title}{{SUMER
  spectral observations of post-flare supra-arcade inflows},} \solphys, 217,
  247, \dodoi{10.1023/B:SOLA.0000006899.12788.22}

\bibitem[{J.~R. Lemen {et~al.}(2012)Lemen, Title, Akin, Boerner, Chou, Drake,
  Duncan, Edwards, Friedlaender, Heyman, Hurlburt, Katz, Kushner, Levay,
  Lindgren, Mathur, McFeaters, Mitchell, Rehse, Schrijver, Springer, Stern,
  Tarbell, Wuelser, Wolfson, Yanari, Bookbinder, Cheimets, Caldwell, Deluca,
  Gates, Golub, Park, Podgorski, Bush, Scherrer, Gummin, Smith, Auker, Jerram,
  Pool, Soufli, Windt, Beardsley, Clapp, Lang, \& Waltham}]{Lemen2012}
Lemen, J.~R., Title, A.~M., Akin, D.~J., {et~al.} 2012, \bibinfo{title}{The
  {{Atmospheric Imaging Assembly}} ({{AIA}}) on the {{Solar Dynamics
  Observatory}} ({{SDO}}),} Solar Physics, 275, 17,
  \dodoi{10.1007/s11207-011-9776-8}

\bibitem[{F. {Lepreti} {et~al.}(2001){Lepreti}, {Carbone}, \&
  {Veltri}}]{lepreti2001solar}
{Lepreti}, F., {Carbone}, V., \& {Veltri}, P. 2001, \bibinfo{title}{{Solar
  Flare Waiting Time Distribution: Varying-Rate Poisson or L{\'e}vy
  Function?},} \apjl, 555, L133, \dodoi{10.1086/323178}

\bibitem[{Z.~F. {Li} {et~al.}(2021){Li}, {Cheng}, {Ding}, {Reeves}, {Kittrell},
  {Weber}, \& {McKenzie}}]{li2021thermodynamic}
{Li}, Z.~F., {Cheng}, X., {Ding}, M.~D., {et~al.} 2021,
  \bibinfo{title}{{Thermodynamic Evolution of Solar Flare Supra-arcade
  Downflows},} \apj, 915, 124, \dodoi{10.3847/1538-4357/ac043e}

\bibitem[{J. {Lin}(2004){Lin}}]{Lin2004}
{Lin}, J. 2004, \bibinfo{title}{{Motions of Flare Ribbons and Loops in Various
  Magnetic Configurations},} \solphys, 222, 115,
  \dodoi{10.1023/B:SOLA.0000036875.14102.39}

\bibitem[{J. {Lin} {et~al.}(2007){Lin}, {Li}, {Forbes}, {Ko}, {Raymond}, \&
  {Vourlidas}}]{Lin2007}
{Lin}, J., {Li}, J., {Forbes}, T.~G., {et~al.} 2007, \bibinfo{title}{{Features
  and Properties of Coronal Mass Ejection/Flare Current Sheets},} \apjl, 658,
  L123, \dodoi{10.1086/515568}

\bibitem[{R. {Liu}(2013){Liu}}]{Liu2013}
{Liu}, R. 2013, \bibinfo{title}{{Dynamical processes at the vertical current
  sheet behind an erupting flux rope},} \mnras, 434, 1309,
  \dodoi{10.1093/mnras/stt1090}

\bibitem[{R. {Liu} \& Y. {Wang}(2021){Liu} \& {Wang}}]{Liu&Wang2021}
{Liu}, R., \& {Wang}, Y. 2021, \bibinfo{title}{{Investigation on the
  spatiotemporal structures of supra-arcade spikes},} \aap, 653, A51,
  \dodoi{10.1051/0004-6361/202140847}

\bibitem[{D.~E. {McKenzie}(2013){McKenzie}}]{mckenzie2013turbulent}
{McKenzie}, D.~E. 2013, \bibinfo{title}{{Turbulent Dynamics in Solar Flare
  Sheet Structures Measured with Local Correlation Tracking},} \apj, 766, 39,
  \dodoi{10.1088/0004-637X/766/1/39}

\bibitem[{D.~E. {McKenzie} \& H.~S. {Hudson}(1999){McKenzie} \&
  {Hudson}}]{mckenzie1999x}
{McKenzie}, D.~E., \& {Hudson}, H.~S. 1999, \bibinfo{title}{{X-Ray Observations
  of Motions and Structure above a Solar Flare Arcade},} \apjl, 519, L93,
  \dodoi{10.1086/312110}

\bibitem[{W.~D. {Pesnell} {et~al.}(2012){Pesnell}, {Thompson}, \&
  {Chamberlin}}]{Pesnell2012}
{Pesnell}, W.~D., {Thompson}, B.~J., \& {Chamberlin}, P.~C. 2012,
  \bibinfo{title}{{The Solar Dynamics Observatory (SDO)},} \solphys, 275, 3,
  \dodoi{10.1007/s11207-011-9841-3}

\bibitem[{K.~K. {Reeves} {et~al.}(2017){Reeves}, {Freed}, {McKenzie}, \&
  {Savage}}]{reeves2017exploration}
{Reeves}, K.~K., {Freed}, M.~S., {McKenzie}, D.~E., \& {Savage}, S.~L. 2017,
  \bibinfo{title}{{An Exploration of Heating Mechanisms in a Supra-arcade
  Plasma Sheet Formed after a Coronal Mass Ejection},} \apj, 836, 55,
  \dodoi{10.3847/1538-4357/836/1/55}

\bibitem[{K.~K. {Reeves} {et~al.}(2019){Reeves}, {T{\"o}r{\"o}k}, {Miki{\'c}},
  {Linker}, \& {Murphy}}]{Reeves2019}
{Reeves}, K.~K., {T{\"o}r{\"o}k}, T., {Miki{\'c}}, Z., {Linker}, J., \&
  {Murphy}, N.~A. 2019, \bibinfo{title}{{Exploring Plasma Heating in the
  Current Sheet Region in a Three-dimensional Coronal Mass Ejection
  Simulation},} \apj, 887, 103, \dodoi{10.3847/1538-4357/ab4ce8}

\bibitem[{T. {Samanta} {et~al.}(2021){Samanta}, {Tian}, {Chen}, {Reeves},
  {Cheung}, {Vourlidas}, \& {Banerjee}}]{samanta2021plasma}
{Samanta}, T., {Tian}, H., {Chen}, B., {et~al.} 2021, \bibinfo{title}{{Plasma
  heating induced by tadpole-like downflows in the flaring solar corona},} The
  Innovation, 2, 100083, \dodoi{10.1016/j.xinn.2021.100083}

\bibitem[{R. {S{\'a}nchez} \& D. {Newman}(2018){S{\'a}nchez} \&
  {Newman}}]{Survival2018}
{S{\'a}nchez}, R., \& {Newman}, D. 2018, {A Primer on Complex Systems}, Vol.
  943, \dodoi{10.1007/978-94-024-1229-1}

\bibitem[{S.~L. {Savage} \& D.~E. {McKenzie}(2011){Savage} \&
  {McKenzie}}]{Savage&McKenzie2011}
{Savage}, S.~L., \& {McKenzie}, D.~E. 2011, \bibinfo{title}{{Quantitative
  Examination of a Large Sample of Supra-arcade Downflows in Eruptive Solar
  Flares},} \apj, 730, 98, \dodoi{10.1088/0004-637X/730/2/98}

\bibitem[{S.~L. {Savage} {et~al.}(2012){Savage}, {McKenzie}, \&
  {Reeves}}]{savage2012re}
{Savage}, S.~L., {McKenzie}, D.~E., \& {Reeves}, K.~K. 2012,
  \bibinfo{title}{{Re-interpretation of Supra-arcade Downflows in Solar
  Flares},} \apjl, 747, L40, \dodoi{10.1088/2041-8205/747/2/L40}

\bibitem[{G. {Schwarz}(1978){Schwarz}}]{BIC1978}
{Schwarz}, G. 1978, \bibinfo{title}{{Estimating the Dimension of a Model},}
  Annals of Statistics, 6, 461

\bibitem[{C. {Shen} {et~al.}(2022){Shen}, {Chen}, {Reeves}, {Yu}, {Polito}, \&
  {Xie}}]{shen2022origin}
{Shen}, C., {Chen}, B., {Reeves}, K.~K., {et~al.} 2022, \bibinfo{title}{{The
  origin of underdense plasma downflows associated with magnetic reconnection
  in solar flares},} Nature Astronomy, 6, 317,
  \dodoi{10.1038/s41550-021-01570-2}

\bibitem[{M. Skender \& G. Lapenta(2010)Skender \&
  Lapenta}]{skender2010instability}
Skender, M., \& Lapenta, G. 2010, \bibinfo{title}{On the instability of a
  quasiequilibrium current sheet and the onset of impulsive bursty
  reconnection,} Physics of plasmas, 17

\bibitem[{N. Sugiura(1978)Sugiura}]{AICc1978}
Sugiura, N. 1978, \bibinfo{title}{Further analysis of the data by akaike's
  information criterion and the finite corrections: Further analysis of the
  data by akaike's,} Communications in Statistics-theory and Methods, 7, 13

\bibitem[{G. {Tan} {et~al.}(2022){Tan}, {Hou}, \& {Tian}}]{tan2022statistical}
{Tan}, G., {Hou}, Y., \& {Tian}, H. 2022, \bibinfo{title}{{Statistical
  investigation of the kinematic and thermal properties of supra-arcade
  downflows observed during a solar flare},} \mnras, 516, 3120,
  \dodoi{10.1093/mnras/stac2470}

\bibitem[{Z. {{\v{S}}vestka} {et~al.}(1998){{\v{S}}vestka}, {F{\'a}rn{\'\i}k},
  {Hudson}, \& {Hick}}]{vsvestka1998large}
{{\v{S}}vestka}, Z., {F{\'a}rn{\'\i}k}, F., {Hudson}, H.~S., \& {Hick}, P.
  1998, \bibinfo{title}{{Large-Scale Active Coronal Phenomena in Yohkoh SXT
  Images IV. Solar Wind Streams from Flaring Active Regions},} \solphys, 182,
  179, \dodoi{10.1023/A:1005033717284}

\bibitem[{M.~S. {Wheatland}(2000){Wheatland}}]{wheatland2000origin}
{Wheatland}, M.~S. 2000, \bibinfo{title}{{The Origin of the Solar Flare
  Waiting-Time Distribution},} \apjl, 536, L109, \dodoi{10.1086/312739}

\bibitem[{W. Willinger {et~al.}(2004)Willinger, Alderson, Doyle, \&
  Li}]{willinger2004}
Willinger, W., Alderson, D., Doyle, J., \& Li, L. 2004, in Proceedings of the
  2004 Winter Simulation Conference, 2004., Vol.~1, 141,
  \dodoi{10.1109/WSC.2004.1371310}

\bibitem[{X. {Xie} {et~al.}(2022){Xie}, {Reeves}, {Shen}, \&
  {Ingram}}]{xie2022statistical}
{Xie}, X., {Reeves}, K.~K., {Shen}, C., \& {Ingram}, J.~D. 2022,
  \bibinfo{title}{{Statistical Study of the Kinetic Features of Supra-arcade
  Downflows Detected from Multiple Solar Flares},} \apj, 933, 15,
  \dodoi{10.3847/1538-4357/ac695d}

\bibitem[{X. {Xie} {et~al.}(2025){Xie}, {Shen}, {Reeves}, {Chen}, {Li}, {Guo},
  {Yu}, {Wei}, \& {Dong}}]{Xie2025}
{Xie}, X., {Shen}, C., {Reeves}, K.~K., {et~al.} 2025,
  \bibinfo{title}{{Anisotropic Turbulent Flows Observed in Above-the-loop-top
  Regions during Solar Flares},} \apjl, 984, L27,
  \dodoi{10.3847/2041-8213/adc91b}

\bibitem[{J. {Xue} {et~al.}(2020){Xue}, {Su}, {Li}, \&
  {Zhao}}]{xue2020thermodynamical}
{Xue}, J., {Su}, Y., {Li}, H., \& {Zhao}, X. 2020,
  \bibinfo{title}{{Thermodynamical Evolution of Supra-arcade Downflows},} \apj,
  898, 88, \dodoi{10.3847/1538-4357/ab9a3d}

\end{thebibliography}
\bibliographystyle{aasjournalv7}

\end{document}